# Fully retarded van der Waals interaction between dielectric nanoclusters

Hye-Young Kim [a,b], Jorge. O. Sofo [a,b], Darrell Velegol [b,c], and Milton. W. Cole [a,b]

*Departments of Physics [a] and Chemical Engineering [c] and the Materials Research Institute [b]*
*The Pennsylvania State University, University Park, PA 16802*

ABSTRACT

The fully retarded dispersion interaction between an atom and a cluster or between two clusters is calculated. Results obtained with two different methods are compared. One is to consider a cluster as a collection of many atoms and evaluate the sum of two-body and three-body interatomic interactions, a common assumption. The other method, valid at large separation, is to consider each cluster as a point particle, characterized by a polarizability tensor, and evaluate the inter-cluster interaction. This method employs the static polarizability, evaluated by including all many-body (MB) intra-cluster atomic interactions self-consistently, which yields the full inter-cluster interaction, including all MB terms. A comparison of the results from the two methods reveals that the contribution of the higher-than-three-body MB interactions is always attractive and non-negligible, with a relative importance that varies with geometry. The procedure is quite general and is applicable to any shape or size of dielectric clusters, in principle. We present numerical results for clusters composed of atoms with polarizability consistent with silica, for which the higher-than-three-body MB correction term can be as high as 42% of the atomic pair-wise sum. The full result is quite sensitive to the anisotropic structure of the cluster, in contrast to the result found in the additive case, which is orientation independent. We also present a power law expansion of the total van der Waals (VDW) interaction as a series of n-body interaction terms.

## I. INTRODUCTION

In many studies of condensed phases of matter, researchers have assumed that the relevant forces between two macroscopic substances arise from pair-wise dispersion interactions between the constituent atoms. (Here, we let the term "atom" refer to any polarizable, small unit such as a molecule.) With that assumption, one can easily compute the net force on a specified atom by summing force contributions from neighboring atoms. Equivalently, the net potential energy is considered as a pairwise sum of interactions between atoms. While these assumptions simplify computations, one may wonder whether they are valid. Indeed, in the nonretarded regime, it is well known that many-body (MB) corrections to additivity are important in many cases, including the cases of the inert gas fluids, where one obtains better agreement with experimental data when the three-body interaction terms are included [1-5]. Researchers have tended to generalize these results to other systems and adopt one, or other, of three strategies: assume that MB corrections to additivity are negligible, employ a two-body potential that nominally includes 3-body interactions or evaluate three-body interactions explicitly and assume that they represent the only significant correction to the additivity approximation.

For more than 50 years, it has been realized that the finite velocity of light affects van der Waals (VDW) interactions: this phenomenon is called *retardation*. Effects of retardation appear when separations between interacting bodies are larger than about 10 nm. These are receiving increasing attention for reasons of both fundamental and applied science [6-13]. The evaluation of these interactions in the *fully retarded* regime, of very large separation, is simpler than in the nonretarded regime, because one needs to know just the static polarizability in the former case, while latter requires the frequency-dependent polarizability. This fully retarded regime is sometimes referred to as the "Casimir force" regime since Casimir and colleagues developed the first treatment of such forces [14, 15].

Up to now, there has not been any detailed study of MB effects in the VDW interaction between clusters in the fully-retarded regime [16]. An efficient way (exact at large separation) to evaluate the interaction is to consider each cluster as a single large molecule with a well-defined polarizability tensor and use the rigorous expression derived for the VDW interaction between molecules. There is a body of experimental and theoretical information concerning the polarizability for the case of metallic clusters [17-19]. Unfortunately, since this is not the case for dielectric clusters, one needs an alternative way of calculating VDW interactions between such clusters. One way to achieve this has been to add two-body interactions between constituent atoms. This traditional way of calculating the VDW interactions between two clusters ignores the interactions between atoms comprising a cluster and results in the effective polarizability of the cluster as a simple sum of the individual atomic polarizabilities.

In our previous work [20], we have evaluated the static polarizability of a cluster of arbitrary shape and size, using a microscopic, self-consistent method which includes all MB interactions between atoms in the cluster. It has been shown that this cluster polarizability differs from the simple sum of the individual atomic polarizabilities. Using these cluster polarizabilities, we can evaluate the VDW interaction between dielectric clusters at large separation, considering each cluster as a single large molecule. Since this interaction includes implicitly all MB interactions, we obtain the full VDW interaction, including all MB terms, in the fully-retarded regime.

The outline of this paper is the following. In the next section, we describe the basic formulation and the set of cluster configurations considered. In section III, we present an evaluation of the fully retarded interaction between various clusters and a single atom. This is followed, in section IV, by an evaluation of the interaction between two clusters. In sections III and IV, the "full" (exact) results are compared with the "partial" (approximate) results based on the addition of the two-body and three-body interactions. Our results (typically, inadequacy of the additivity approximation) are summarized and discussed in section V.

## II. RETARDED DISPERSION POTENTIAL : FORMULATION

Here, we calculate the inter-cluster interaction by treating each cluster as a point particle, with a calculated static polarizability [20], and compare this result with that derived by adding two-body interactions between atoms belonging to different clusters. Since a cluster, in general, has an anisotropic polarizability, we employ the expression for the two-body retarded dispersion interaction between anisotropic particles [21, 22] for the inter-cluster interaction:

$$V_{aniso}^{(2)} = -\frac{\hbar c}{8\pi} \frac{\aleph(A:B)}{r_{AB}^{7}} \tag{1}$$

where

$$\aleph(A:B) = 13 \left[ \alpha_{11}^{(A)} \alpha_{11}^{(B)} + \alpha_{22}^{(A)} \alpha_{22}^{(B)} \right] + 20 \; \alpha_{33}^{(A)} \alpha_{33}^{(B)} + 26 \; \alpha_{12}^{(A)} \alpha_{12}^{(B)} \\ - 30 \left[ \alpha_{23}^{(A)} \alpha_{23}^{(B)} + \alpha_{31}^{(A)} \alpha_{31}^{(B)} \right] \tag{2}$$

Here, the index 3 indicates the direction along the line connecting the two clusters and the indices 1 and 2 refer to the perpendicular directions. Symmetry can simplify this expression. For example, $\aleph(A:B) = 46 \alpha_{atom}^2$ when A and B represent two identical atoms with isotropic atomic polarizability $\alpha_{atom}$. Also, when an atom on the z-axis interacts with a cluster (A) on the z-axis possessing rotational symmetry about the z-axis, the expression becomes

$$\aleph(A:B) = 2\alpha_{atom} \left[ 13\alpha_{\perp}^{(A)} + 10\alpha_{//}^{(A)} \right] \tag{3}$$

where, $\alpha_{\perp}^{(A)} = \alpha_{11}^{(A)} = \alpha_{22}^{(A)}$ and $\alpha_{//}^{(A)} = \alpha_{33}^{(A)}$. In recent work [20], we evaluated the static polarizabilities of clusters of various shapes and sizes by utilizing a microscopic, self-consistent, linear response method, which is exact within the linear and dipolar approximations. We will use these polarizability values to calculate the two-body retarded dispersion interaction between a cluster and an atom and also between two clusters. Since this polarizability includes all intra-cluster MB terms, the resulting interaction includes all MB contributions.

For comparison, we evaluate the inter-cluster VDW interaction by considering each cluster as a composite of atoms and adding pairwise interatomic interactions. Here, each "atom" is characterized with an isotropic atomic polarizability, as described below. We also evaluate the leading non-additive correction term by summing the three-body atomic interactions. The three-body dispersion interaction, here, involves three isotropic atoms A, B and C, where A and B are nearby atoms within a cluster (i.e., in nonretarded separation regime) while C is far from those two (i.e., in retarded regime), located in the other cluster, as shown in Fig.1. Therefore, there is retardation in the interactions A-C and B-C, but not A-B. An analytic formula for such a three-body interaction is derived following the procedure of Aub and Zienau [22] ; the detailed derivation and resulting expression are provided in Appendix I.  As in the Axilrod-Teller-Muto (ATM) [23, 24] expression for the non-retarded three-body interaction, this partially retarded three-body interaction depends on the inner angles of the triangle connecting the three atoms involved, $\theta_A$, $\theta_B$ and $\theta_C$. Note that the distance dependence of the "partially retarded" three-body interaction is $r^{-10}$, in contrast to the $r^{-9}$ dependence of the non-retarded three-body (ATM) interaction.

To explore the dependence on the cluster geometry, several shapes, sizes and orientations are investigated: linear clusters (with dimension $1 \times 1 \times L$) with $L = 2$ to $1000$, square prism clusters ($2 \times 2 \times L$) with $L = 2$ to $200$, cubic clusters ($L \times L \times L$) with $L = 2$ to $10$ and square monolayer clusters ($1 \times L \times L$) with $L = 2$ to $30$. Here, each cluster is composed of atoms residing at simple cubic lattice sites with lattice constant $a_0$, and the lattice constant is the unit of length. For quantitative estimates, we use atomic polarizabilities and lattice constants of silica [20, 25]. The "atomic" polarizability ($\alpha_{atom}$) is determined from the Clausius-Mossotti relation, using known dielectric spectra of fused silica ($SiO_2$) [20, 26], and the lattice constant is obtained from the known number density ($n_s$); these yield $\alpha_{atom} = 3.76 \, \overset{\circ}{\text{A}}{}^3$, $n_s = 0.0334 \, \overset{\circ}{\text{A}}{}^{-3}$ and $a_0 = 3.569 \, \overset{\circ}{\text{A}}$ for silica. Although results are presented only for silica clusters, the results for clusters composed of other substances would differ only quantitatively, through the products $n_s \alpha_{atom}$ [20], as long as the cluster geometry remains the same.

## III. FULLY RETARDED ATOM-CLUSTER INTERACTIONS

We compare the full inter-cluster VDW interaction $V$ with $V^{(2)}$ computed by summing two-body atomic retarded interactions using Eq. (1) with $\aleph = 46 \alpha_{atom}^2$ and $V^{(3)}$ computed by summing three-body atomic partially retarded interactions using Eq. (17). In the following, instead of concentrating on the individual values, $V$, $V^{(2)}$ and $V^{(3)}$, we will focus on the dimensionless quantity

$$R_m = \frac{\left(V - V^{(2)}\right)}{V^{(2)}} \frac{1}{n_s \alpha_{atom}} \tag{4}$$

and

$$R_3 = \frac{V^{(3)}}{V^{(2)}} \frac{1}{n_s \alpha_{atom}} \qquad . \tag{5}$$

These are the ratios of the MB and three-body atomic interaction contributions to the pair-wise sum of two-body atomic interaction, respectively, divided by the product $n_s \alpha_{atom}$ [27]. The difference between these ratios represents the contribution of many-body interactions of higher order than three-body interactions:

$$R_{4h} = R_m - R_3 \tag{6}$$

The subscript $4h$ means that fourth and higher order contributions are included.

Due to the polarizability anisotropy of the cluster, the interaction depends on the orientation of the cluster; the results of $R_{4h}$ for various clusters are shown in Figs. 2 to 6 as a function of the total number of atoms ($N$) in each cluster. In Fig. 2, although the magnitude of the higher-order many-body contribution ($R_{4h}$) is a function of both shape and orientation, we find that all of these higher order terms represent attractive contributions to the potential. In contrast, as seen in Figs. 3 to 6, the three-body contributions may be either attractive or repulsive, depending on the orientation [25, 28]. The magnitude of these higher-order contributions vary between 5 % of the two-body energy for isotropic cubic clusters and 42 % for anisotropic square monolayer clusters. Note that the orientation dependence of $R_{4h}$ remains significant even at such large separation between an atom and a cluster except for the case between a square monolayer cluster and an atom. The case of a square monolayer is an exception in that $R_{4h}$ values for parallel and perpendicular orientations coincide. The notable orientation dependences for both linear and square prism clusters are: (i) $R_{4h}$ is nearly 50% larger for parallel orientation than for perpendicular orientation, and (ii) $R_3$ is positive when long dimensions are oriented parallel to the inter-cluster connecting line and is negative for perpendicular orientations. The orientation dependence of (ii) is consistent with that found in the non-retarded three-body interaction [27]. We also note that $R_{4h} > |R_3|$ for all cases studied. Thus the inclusion of just the three-body interactions is a significant underestimates of the many-body interactions. Indeed, in some cases the sign of the three-body term (only) is "wrong", so that the exact result is closer to the two-body energy than to the sum of two-body and three-body energies.

Now, we would like to show that the total atom-cluster interaction can be expanded in a power law series in terms of $n_s \alpha_{atom}$ in which each term corresponds to the n-body atomic terms in the perturbation expansion, where n=2, 3, 4 … . The details of the derivation are as follows. In Ref. [20], in addition to the numerical values of the static polarizability for finite-size clusters, analytic expressions for the static cluster polarizability are derived for both infinite-size clusters and ellipsoidal continuum clusters. In comparing results for linear and square monolayer cluster, excellent agreement was found between polarizabilities of a large linear cluster and of an infinite linear chain, and also between those of a large square monolayer cluster and of a continuum disc. Therefore, we employ the analytic expression for the cluster polarizability of the infinite linear chain and the continuum disc for finite-size linear clusters and square monolayer clusters, respectively, and expand the analytic expression in $n_s \alpha_{atom}$. For example, the fully retarded interaction between an atom and a linear cluster (A) lying along x-axis, perpendicular to the connecting line (z-axis) is:

$$
\begin{aligned}
V &= -\frac{\hbar c}{8\pi} \frac{\alpha_{atom}}{r^7} \left[ 13 \left( \alpha_{xx}^{(A)} + \alpha_{yy}^{(A)} \right) + 20\, \alpha_{zz}^{(A)} \right] \\
&= -\frac{\hbar c}{8\pi} \frac{N\alpha_{atom}^2}{r^7} \left[ 13 f_{//} + 33 f_{\perp} \right]
\end{aligned}
\tag{7}
$$

here, $f_{//} \left( \equiv \dfrac{\alpha_{xx}^{(A)}}{N\alpha_{atom}} \right)$ and $f_{\perp} \left( \equiv \dfrac{\alpha_{yy}^{(A)}}{N\alpha_{atom}} = \dfrac{\alpha_{zz}^{(A)}}{N\alpha_{atom}} \right)$ are the enhancement factors of the cluster polarizability [20]. Analytic expressions for the infinite-size linear chain cluster are $f_{//} = \left[ 1 - k_{//} n_s \alpha_{atom} \right]^{-1}$ and $f_{\perp} = \left[ 1 + k_{\perp} n_s \alpha_{atom} \right]^{-1}$ with $k_{\perp} = k_{//} / 2 = 2\zeta(3) \approx 2.4$. From the reported values of the enhancement factors, $f_{//}$ and $f_{\perp}$, for various sizes of linear cluster [20], one may obtain the corresponding values of $k_{//}$ and $k_{\perp}$, respectively, from the above relations. Note that at a very large separation as in the fully retarded regime, the pairwise summation of atomic interaction equals N times the atomic pair interaction,

$$
V^{(2)} = -\frac{23\hbar c}{4\pi} \frac{N\alpha_{atom}^2}{r^7} \qquad .
\tag{8}
$$

From Eqs. (7) and (8), the dimensionless ratio of interest becomes :

$$
[R_m]_x = \frac{1}{46\, n_s \alpha_{atom}} \left( \frac{33}{1 + k_{\perp} n_s \alpha_{atom}} + \frac{13}{1 - k_{//} n_s \alpha_{atom}} - 46 \right) \quad .
\tag{9}
$$

Here, one may expand this analytic expression for small value of $n_s \alpha_{atom}$ to obtain :

$$[R_m]_x = \frac{1}{46} \sum_{n=3}^{\infty} \left[ \left( 13\, k_{//}^{n-2} + (-1)^n \times 33\, k_{\perp}^{n-2} \right) \left( n_s \alpha_{atom} \right)^{n-3} \right] \qquad . \qquad (10)$$

The term with $n$ in the summation in Eq. (10) corresponds to the n-body atomic interaction term. Values for silica of the first (3-body) term are $-0.152$ for a dimer $(2 \times 1 \times 1)$, $-0.272$ for a decamer $(10 \times 1 \times 1)$, and $-0.365$ for both large $(1000 \times 1 \times 1)$ and infinite $(\infty \times 1 \times 1)$ linear clusters. The negative sign indicates that the three-body interaction is repulsive for this orientation, which is also found for the nonretarded three-body interaction [27]. A similar expansion for the fully retarded interaction between an atom and a linear cluster oriented parallel to the connecting line (z axis) is:

$$[R_m]_z = \frac{1}{23} \sum_{n=3}^{\infty} \left[ \left( 10\, k_{//}^{n-2} + (-1)^n \times 13\, k_{\perp}^{n-2} \right) \left( n_s \alpha_{atom} \right)^{n-3} \right] \qquad . \qquad (11)$$

In this orientation, the three-body term for a silica linear cluster is $0.304$ for a dimer $(1 \times 1 \times 2)$, $0.692$ for a decamer $(1 \times 1 \times 10)$, and $0.730$ for both large $(1000 \times 1 \times 1)$ and infinite $(1 \times 1 \times \infty)$ linear clusters. The positive sign indicates an attractive three-body interaction for this orientation, consistent again with that found for the nonretarded three-body interaction [27]. The results of the expanded VDW interactions between an atom and a linear cluster are shown in Fig. 7. We have also expanded the fully retarded vdW interaction, as above, for a square monolayer cluster, and the results are shown in Fig. 8. The power expansion expressions of $R_m$ for various orientations are listed in the Appendix II. For the square monolayer cluster, the n-body terms involving odd (even) number of atoms ($n$) are repulsive (attractive), except for the 3-body term when the cluster lies along the connecting line. This is an interesting contrast from the results for the linear cluster, where all of the n-body terms are attractive, with the exception of the 3-body term when the cluster lies perpendicular to the connecting line. The results obtained from the direct sum of three-body atomic interactions using eq. (17) are also plotted in Figs. 7 and 8, and they show quite good agreement with those found in the series expansion convincing us that the series expansion indeed gives the MB terms.

For the interaction of an atom with an asymmetric cluster, the total dispersion interaction is always found to be most strongly attractive when the cluster is oriented parallel to the line connecting it to the atom. This finding is also true within the approximation including just pair and three-body interactions. However, the results in Fig. 2 show that the magnitude of higher-order many-body contribution is not negligible. The most dramatic case is the most symmetrical situation, a cubic cluster depicted in Fig. 6. Here, the three-body interaction is essentially zero while the many-body contribution is not negligible (although it is smaller than for asymmetric clusters). Here, note that the vanishing three-body interaction for symmetric clusters has also been found in the non-retarded three-body interaction, at large separation [27]. Note that the series expansion of the expression for a continuum sphere, which corresponds the large size limit of a cubic cluster, gives all the MB contribution terms as zero. We note that the series expansion of the VDW interaction is an asymptotic expression and the accuracy is expected to increase as the cluster size increases.

## IV. FULLY RETARDED CLUSTER-CLUSTER INTERACTIONS

In this section, the interaction between two clusters is also calculated in two ways. We obtain the interactions $V$, $V^{(2)}$, and $V^{(3)}$, the ratios of the many-body ($R_m$) and three-body ($R_3$) interaction contributions to the pair-wise sum of two-body interaction, respectively, and the difference between these two ratios ($R_{4h}$). One can evaluate $V$ for an arbitrary pair of clusters. For interactions between two identical cubic clusters, the MB contribution is small, due to symmetry; since for $N > 500$ $R_{4h} \sim R_m \sim 0.85$, the three-body contribution is negligible while the higher order MB contribution to the two-body contribution is $\sim 10\%$ [see Fig. 9].

For a pair of linear clusters, $V$ depends on three angles: the polar angles ($\theta$ and $\theta'$) and the difference $\Delta\phi$ between their azimuthal angles, where the $z$ axis lies along the inter-cluster vector. We have considered four different relative orientations: $(\theta, \theta', \Delta\phi) = (0, 0, 0)$, $(\pi/2, \pi/2, 0)$, $(0, \pi/2, 0)$, and $(\pi/2, \pi/2, \pi/2)$, denoted *zz*, *xx*, *zx*, and *yx*, respectively. For interactions between two identical square prisms (or lines), the three-body contribution is the most attractive in the zz-configuration and is weakly attractive in the zx-configuration. Figures 10 and 11 show instances when the three-body interaction is repulsive, xx- and yx-configurations. Therefore, if one were to calculate the interaction by summing the two-body and the three-body contributions, the attraction would be the strongest for zz and then case zx, followed by xx and yx. However, if one includes higher-order contributions, the order of the strength of the attraction changes. From strongest to weakest attraction, the order becomes zz, xx, zx, and yx.

Similar changes are also observed in the interaction between two identical square monolayers. For a pair of square monolayer clusters, due to symmetry, *V* also depends on three angles as for linear clusters, except here that the polar angles and the azimuthal angles are defined by the direction of the surface normal vector pointing vertically outwards on one side of the monolayer. We have considered three configurations: $(\theta, \theta', \Delta\phi) = (0, 0, 0)$, $(\pi/2, \pi/2, 0)$, and $(0, \pi/2, 0)$, denoted *xx*, *zz*, and *xz*, respectively, named from the direction of the longer dimension of the monolayer (see Fig. 12). The three-body contribution is strongly repulsive in the xx-configuration, weakly repulsive in the xz-configuration, but *attractive* in the zz-configuration. Hence, if one were to calculate the interaction by summing the two-body and the three-body contributions, the attractive strengths would decrease in the sequence zz, xz and xx. The order changes to zz, xx and xz, however, when we include higher-order MB contributions. It is interesting to observe in Fig. 12 that the magnitude of the higher-than-three-body contributions for xx and zz are almost the same and the difference between the two cases is primarily due to the different three-body contribution.

The full VDW interactions between two clusters are expanded in a power series of $n_s \alpha_{atom}$, as done in the previous section. The analytic expressions are listed in the Appendix II for different shapes and relative orientations of two clusters. The numerical results are shown in Figs. 13 and 14 for silica decamers (N=10) and square monolayer clusters (N=100), respectively. For a pair of decamers, $V^{(2)}$ accounts for only 49% of total VDW interaction for $zz$ configuration and 64%, 75%, and 84% for $xx$, $zx$, and $yx$ configurations, respectively. Similarly, for a pair of square monolayers (N=100), $V^{(2)}$ accounts for 49% of total VDW interaction for $zz$ configuration and 60% and 69% for $xx$ and $xz$ configurations, respectively. While each MB term higher than third order contributes an additional attraction for decamers, these terms for square monolayers alternatively change sign depending on the number of atoms involved in the many-body term (n), being even and odd, respectively. In contrast, the three-body term can be either repulsive or attractive, depending on the relative orientation of the clusters, and the sign agrees with that of the nonretarded 3-body energy [25]. The results obtained from the sum of two-body and three-body atomic interactions, using eqs. (1) and (17), are also plotted in Figs. 13 and 14, and they show good agreement with those found in the series expansion.

## V. CONCLUSION

The fully retarded dispersion interaction between various kinds of clusters with an atom and also with another identical cluster was evaluated utilizing two methods. One is exact, in principle, considering each cluster as a single particle identified with a constant polarizability tensor (as calculated in Ref. 20 including all the many-body interaction terms) and using this to calculate the inter-cluster two-body dispersion interaction. Another is to consider each cluster as a composite of many atoms and add two-body and three-body atomic interactions between the constituent atoms, as often assumed. The comparison between the results from these two methods shows that the contribution of higher-than-three-body dispersion interactions is always attractive and not negligible. Including these higher order MB terms causes a change of the orientation dependence of the interaction between two nanoclusters, relative to that expected when just three-body terms are included as nonadditivity correction. It is also noted that the dependence on the orientation of asymmetric clusters remains significant even at the very large separation of fully retarded regime of interaction. These results demonstrate that the inclusion of the many-body dispersion interactions is important in evaluation of the VDW interactions among nanoclusters. The altered ordering of interaction strengths originates from the strong anisotropy of the cluster polarizability. We have also presented the power law expansion of the total VDW interaction in a series of n-body terms.

## VI. ACKNOWLEDGMENTS

This research is supported by the National Science Foundation through NSF NER Grant No. CTS-0403646 and Ben Franklin Technology Center of Excellence in Nanoparticulate Science and Engineering. We are grateful to Craig Bohren, Lou Bruch, Amand Lucas, Jerry Mahan, and Gautam Mukhopadhyay for helpful discussions.

## APPENDIX I : Three-body atomic dispersion interaction when only one atom is remotely located in fully retarded regime.

Aub and Zienau [22] have derived a general form of the three-body dispersion interaction, which is valid for any separation. The special case of a three-body interaction in our system (see Fig. 1) involves three atoms A, B and C, where A and B are nearby, within a cluster (i.e., nonretarded regime) and C is far from those two (i.e., in retarded regime), located in the other cluster. Therefore, retardation affects the C-A and B-C interactions, but not the A-B interaction. We start with Eqs. (30) and (31) in Ref. [22],

$$V^{(3)} = \frac{\alpha^{(A)}\alpha^{(B)}\alpha^{(C)}\hbar c}{\pi} \int_0^\infty u^6 tr\left\{\widetilde{C}^{(A)}\widetilde{C}^{(B)}\widetilde{C}^{(C)}\right\}$$

(12)

and let the large separation in B-C and C-A as $R_A$ and $R_B$, respectively, and the small separation in A-B as $r_C$ to obtain the tensors,

$$\widetilde{C}_{ij}^{(A)} = \frac{e^{-uR_A}}{u^2 R_A^3}\left[\left(\delta_{ij} - 3\hat{x}_i^{(A)}\hat{x}_j^{(A)}\right)\left(1 + uR_A + u^2 R_A^2\right) + 2\hat{x}_i^{(A)}\hat{x}_j^{(A)}u^2 R_A^2\right]$$

(13)

$$\widetilde{C}_{ij}^{(B)} = \frac{e^{-uR_B}}{u^2 R_B^3}\left[\left(\delta_{ij} - 3\hat{x}_i^{(B)}\hat{x}_j^{(B)}\right)\left(1 + uR_B + u^2 R_B^2\right) + 2\hat{x}_i^{(B)}\hat{x}_j^{(B)}u^2 R_B^2\right]$$

(14)

$$\widetilde{C}_{ij}^{(C)} \sim \frac{e^{-ur_C}}{u^2 r_C^3}\left(\delta_{ij} - 3\hat{x}_i^{(C)}\hat{x}_j^{(C)}\right).$$

(15)

One then substitute eqs. (13-15) into eq. (12) and carry out the trace, utilizing the integration,

$$\int_0^\infty e^{-au}u^n du = \frac{n!}{a^{n+1}} \quad .$$

(16)

Then, one obtains, the final expression:

$$V_{iso}^{(3)} = \frac{2\hbar c}{\pi}\frac{\alpha^{(A)}\alpha^{(B)}\alpha^{(C)}}{R_A^3 R_B^3 r_C^3 (R_A + R_B)}\left[n_I - n_{II}\cos\theta_A\cos\theta_B\cos\theta_C\right.$$
$$\left. - n_{III}(A)\cos^2\theta_A - n_{III}(B)\cos^2\theta_B - n_{III}(C)\cos^2\theta_C\right]$$

(17)

where

$$n_I = 3 + \frac{5R_A^2 + 5R_B^2 + 18R_A R_B}{(R_A + R_B)^2} + \frac{12R_A^2 R_B^2}{(R_A + R_B)^4}$$

(18)

$$n_{II} = \frac{36R_A^2 R_B^2}{(R_A + R_B)^4}$$

(19)

$$n_{III}(A) = \frac{6R_B^2(4R_A + R_B)}{(R_A + R_B)^3}$$

(20)

$$n_{III}(B) = \frac{6R_A^2(R_A + 4R_B)}{(R_A + R_B)^3} \tag{21}$$

$$n_{III}(C) = n_{III}(A) + n_{III}(B) + \frac{24R_A^2 R_B^2}{(R_A + R_B)^4} \tag{22}$$

where

$$\hat{x}_B \cdot \hat{x}_C = -\cos\theta_A, \tag{23}$$

$$R_A \gg r_C \text{ and } R_B \gg r_C \qquad . \tag{24}$$

## APPENDIX II : Many-body power series expansion of a full VDW interaction

The polarizability enhancement factors [20] are $f_\perp = \dfrac{1}{1 + k_\perp n_s \alpha_{atom}}$ and $f_{//} = \dfrac{1}{1 - k_{//} n_s \alpha_{atom}}$ for both an infinite line ($\infty \times 1 \times 1$) and a continuum disc cluster. Here, $k_\perp = \dfrac{k_{//}}{2} = 2\varsigma(3) \approx 2.40$ for an infinite line ($\infty \times 1 \times 1$) and $k_{//} = \dfrac{k_\perp}{2} = \dfrac{4\pi}{3} \approx 4.19$ for a continuum disc, where $\varsigma(3) \approx 1.20205$ is

the Riemann zeta function [20, 29]. Also, for finite-size clusters [20], $k_{//} = 4.27$ and $k_\perp = 2.06$

for a decamer, $k_\perp = k_{//}/2 = 1.0$ for a dimer, and $k_{//} = 3.40$ and $k_\perp = 6.36$ for a square monolayer

(N=100).

1. An atom and a linear cluster

    a. VDW interaction between an atom and a linear cluster oriented perpendicular to the

connecting line (z axis): $(\theta, \phi) = (\pi/2, 0)$

$$[R_m]_x = \frac{1}{46} \sum_{n=3}^{\infty} \left[ \left( 13\, k_{//}^{n-2} + (-1)^n \times 33\, k_{\perp}^{n-2} \right) \left( n_s \alpha_{atom} \right)^{n-3} \right]$$ (25)

b. VDW interaction between an atom and a linear cluster oriented parallel to the connecting line (z axis): $(\theta, \phi) = (0, 0)$

$$[R_m]_z = \frac{1}{23} \sum_{n=3}^{\infty} \left[ \left( 10\, k_{//}^{n-2} + (-1)^n \times 13\, k_{\perp}^{n-2} \right) \left( n_s \alpha_{atom} \right)^{n-3} \right]$$ (26)

2. An atom and a square monolayer cluster

a. VDW interaction between an atom and a square monolayer lying perpendicular to the connecting line (z axis): $(\theta, \phi) = (0, 0)$

$$[R_m]_x = \frac{1}{23} \sum_{n=3}^{\infty} \left[ \left( 13\, k_{//}^{n-2} + (-1)^n \times 10\, k_{\perp}^{n-2} \right) \left( n_s \alpha_{atom} \right)^{n-3} \right]$$ (27)

b. VDW interaction between an atom and a square monolayer lying parallel to the connecting line (z axis): $(\theta, \phi) = (\pi/2, 0)$

$$[R_m]_z = \frac{1}{46} \sum_{n=3}^{\infty} \left[ \left( 33\, k_{//}^{n-2} + (-1)^n \times 13\, k_{\perp}^{n-2} \right) \left( n_s \alpha_{atom} \right)^{n-3} \right]$$ (28)

3. Two linear clusters

a. VDW interaction between two linear clusters perpendicular to the connecting line (xx configuration); $(\theta, \theta', \Delta\phi) = (\pi/2, \pi/2, 0)$:

$$[R_m]_{xx} = \frac{1}{46} \sum_{n=3}^{\infty} \left[ (n-1) \left( 13\, k_{//}^{n-2} + (-1)^n \times 33\, k_{\perp}^{n-2} \right) \left( n_s \alpha_{atom} \right)^{n-3} \right]$$ (29)

b. VDW interaction between two linear clusters, where both are parallel to the connecting line (zz configuration); $(\theta, \theta', \Delta\phi) = (0, 0, 0)$:

$$[R_m]_{zz} = \frac{1}{23} \sum_{n=3}^{\infty} \left[ (n-1) \left( 10\, k_{//}^{n-2} + (-1)^n \times 13\, k_{\perp}^{n-2} \right) \left( n_s \alpha_{atom} \right)^{n-3} \right]$$ (30)

c. VDW interaction between two linear clusters, where one is parallel to and one is perpendicular to the connecting line (zx configuration); $(\theta, \theta', \Delta\phi) = (0, \pi/2, 0)$:

$$[R_m]_{zz} = \frac{1}{46} \sum_{n=3}^{\infty} \left[ \left( 33 \frac{\left(k_{//}^{n-1} - k_{\perp}^{n-1}\right)}{\left(k_{//} + k_{\perp}\right)} + (-1)^n \times 13\,(n-1)\,k_{\perp}^{n-2} \right) \left(n_s \alpha_{atom}\right)^{n-3} \right] \tag{31}$$

d. VDW interaction between two linear clusters, where one is parallel to and one is perpendicular to the connecting line (yx configuration); $(\theta, \theta', \Delta\phi) = (\pi/2, \pi/2, \pi/2)$:

$$[R_m]_{zz} = \frac{1}{23} \sum_{n=3}^{\infty} \left[ \left( 13 \frac{\left(k_{//}^{n-1} - k_{\perp}^{n-1}\right)}{\left(k_{//} + k_{\perp}\right)} + (-1)^n \times 10\,(n-1)\,k_{\perp}^{n-2} \right) \left(n_s \alpha_{atom}\right)^{n-3} \right] \tag{32}$$

4. Two square monolayer clusters

a. VDW interaction between two square monolayer clusters perpendicular to the connecting line (xx configuration); $(\theta, \theta', \Delta\phi) = (0, 0, 0)$:

$$[R_m]_{xx} = \frac{1}{23} \sum_{n=3}^{\infty} \left[ (n-1)\left( 13\,k_{//}^{n-2} + (-1)^n \times 10\,k_{\perp}^{n-2} \right) \left(n_s \alpha_{atom}\right)^{n-3} \right] \tag{33}$$

b. VDW interaction between two square monolayer clusters, where both are parallel to the connecting line (zz configuration); $(\theta, \theta', \Delta\phi) = (\pi/2, \pi/2, 0)$:

$$[R_m]_{zz} = \frac{1}{46} \sum_{n=3}^{\infty} \left[ (n-1)\left( 33\,k_{//}^{n-2} + (-1)^n \times 13\,k_{\perp}^{n-2} \right) \left(n_s \alpha_{atom}\right)^{n-3} \right] \tag{34}$$

c. VDW interaction between two square monolayer clusters, where one is parallel to and one is perpendicular to the connecting line (zx configuration); $(\theta, \theta', \Delta\phi) = (0, \pi/2, 0)$:

$$[R_m]_{zz} = \frac{1}{46} \sum_{n=3}^{\infty} \left[ \left( 33 \frac{\left(k_{//}^{n-1} - k_{\perp}^{n-1}\right)}{\left(k_{//} + k_{\perp}\right)} + 13\,(n-1)\,k_{//}^{n-2} \right) \left(n_s \alpha_{atom}\right)^{n-3} \right] \tag{35}$$

**REFERENCES**

[1] L. W. Bruch, M. W. Cole, and E. Zaremba, *Physical Adsorption: Forces and Phenomena* (Oxford University Press, Oxford, 1997).

[2] H. T. Davis, *Statistical Mechanics of Phases, Interfaces and Thin Films* (VCH, New York, 1996).

[3] John A. Barker,  in *Simple molecular systems at very  high density*, ed. A. Polian, P. Loubeyre and N. Boccara (Plenum, New York, 1989), pp.331-351.

[4] K. Szalewicz, R. Bukowski, and B. Jeziorski, in *Theory and Applications of Computational Chemistry: The First Forty Years*, ed. C. Dykstra et al. (Elsevier B. V., 2005), Ch. 33.

[5] J.A. Barker, R.O. Watts, J.K. Lee, T.P. Schaefer and Y.T.Lee, Mol. Phys. 21, 657 (1974).

[6] V. A. Parsegian, *Van der Waals Forces*, (Cambridge U. P., 2005).

[7] C. I. Sukenik, M. G. Boshier, D. Cho, V. Sandoghdar, and E. A. Hinds, *Phys. Rev. Lett.* 70, 560 (1993).

[8] U. Mohideen and A. Roy, *Phys. Rev. Lett.* 81, 4549 (1998).

[9] Y. Imry, *Phys. Rev. Lett.* **95**, 080404 (2005).

[10] V. Mkrtchian, V. A. Parsegian, R. Podgornik, and W. M. Saslow, *Phys. Rev. Lett.* **91**, 220801 (2003).

[11] M. Kardar and R. Golestanian, *Rev. Mod. Phys.* **71**, 1233 (1999).

[12] S. K. Lamoreaux, *Phys. Rev. Lett.* **78**, 5 (1997)

[13] R. Garcia and M. H. W. Chan, *Phys. Rev. Lett.* **83**, 1187 (1999).

[14] H. B. G. Casimir and D. Polder, *Phys. Rev.* **73**, 360 (1948).

[15] H. B. G. Casimir, *Proc. K. Ned. Akad. Wet.* **60**, 793 (1948).

[16] The only study known to us, M.M.Calbi, S.M.Gatica,D.Velegol and M.W.Cole, Physical Review A 67, 033201 (2003); was limited to the case of spherical Na clusters.

[17] M. Manninen, R. M. Nieminen, and M. J. Puska, *Phys. Rev.* **B33**, 4289 (1986).

[18] Arup Banerjee and Manoj K. Harbola, *J. Chem. Phys.* **117**, 7845 (2002).

[19] W. D. Knight, K. Clemenger, W. A. de Heer, and W. A. Saunders, *Phys. Rev.* **B31**, R2539 (1985).

[20] H.-Y. Kim, J. O. Sofo, D. Velegol, M. W. Cole, and G. Mukhopadhyay, *Phys. Rev.* A72, 053201 (2005).

[21] E. A. Power and T. Thirunamachandran, *Proc. R. Soc. Lond.* **A401**, 267 (1985).

[22] M. R. Aub and S. Zienau, *Proc. R. Soc. Lond.* **A257**, 464 (1960).

[23] B. M. Axilrod and E. Teller, *J. Chem. Phys.* **11**, 299 (1943).

[24] Y. Muto, *Proceedings of the Physico-Mathematical Society of Japan* **17**, 629 (1943).

[25] S. M. Gatica, M. W. Cole and D. Velegol, *Nano Lett.* **5**, 169 (2005).

[26] R. J. Hunter, *Foundations of Colloid Science*, vol. 1, Section 4.6 (Oxford, New York, 1986).

[27] S. M. Gatica, M. M. Calbi, M. W. Cole, and D. Velegol, *Phys. Rev.* **B68**, 205409 (2003).

[28] H.-Y. Kim and M. W. Cole, *Phys. Rev.* **B35**, 3990 (1987).

[29] *Handbook of Mathematical Functions*, edited by M. Abramowitz and I. A. Stegun (Dover, New York, 1972), p.807.

**FIGURE 1.** (Color online) Configuration of three-body atomic interaction formulation.

**FIGURE 2.** (Color online) The relative (to two-body) many-body dispersion interaction of higher order than the three-body term between one atom and various clusters: cubic cluster (green filled circle), linear cluster(red filled) and square prism (blue open) (triangle and diamond for orientations with long dimension either parallel or perpendicular to the connecting line, respectively), and square monolayer cluster (cross and open square for orientations with long dimension parallel and perpendicular to the connecting line, respectively). Solid and dotted curves are drawn to guide eyes for parallel and perpendicular orientations, respectively.

**FIGURE 3.** (Color online) Ratios $R_m$, $R_3$ and $R_{4h}$ for dispersion interaction between an atom and a square monolayer cluster: $R_m$ (red filled square), $R_3$ (blue open square) and $R_{4h}$ (black cross) for orientation with planar surface parallel to the connecting line (z-configuration), and $R_m$ (red filled circle), $R_3$ (blue open circle) and $R_{4h}$ (black open diamond) for orientation perpendicular to the connecting line (x-configuration). Solid and dotted curves are drawn to guide eyes for parallel and perpendicular orientations, respectively.

**FIGURE 4.** (Color online) Ratios $R_m$, $R_3$ and $R_{4h}$ for dispersion interaction between an atom and a linear cluster: $R_m$ (red filled square), $R_3$ (blue open square) and $R_{4h}$ (black cross) for orientation with long dimension parallel to the connecting line (z-configuration), and $R_m$ (red filled circle), $R_3$ (blue open circle) and $R_{4h}$ (black open diamond) for orientation perpendicular to the connecting line (x-configuration). Solid and dotted curves are drawn to guide eyes for parallel and perpendicular orientations, respectively.

**FIGURE 5.** (Color online) Ratios $R_m$, $R_3$ and $R_{4h}$ for dispersion interaction between an atom and a square prism cluster: $R_m$ (red filled square), $R_3$ (blue open square) and $R_{4h}$ (black cross)

for orientation with long dimension parallel to the connecting line (z-configuration), and $R_m$ (red filled circle), $R_3$ (blue open circle) and $R_{4h}$ (black open diamond) for orientation perpendicular to the connecting line (x-configuration). Solid and dotted curves are drawn to guide eyes for z- and x-configurations, respectively.

**FIGURE 6.** (Color online) Ratios $R_m$, $R_3$ and $R_{4h}$ for dispersion interaction between an atom and a cubic cluster as shown in the inset: $R_m$ (red open square), $R_3$ (blue filled square) and $R_{4h}$ (black cross). The data of $R_m$ and $R_{4h}$ coincide within the resolution of the figure.

**FIGURE 7.** (Color online) Fraction of total VDW interaction energy between a decamer and an atom, as a function of the highest order of many-body terms included. Triangles (circles) correspond to the z (x) -configuration of the decamer. Open symbols are from direct calculation of 2-body and 3-body sum. Dotted curves are drawn to guide eyes.

**FIGURE 8.** (Color online) Fraction of total VDW interaction energy between a square monolayer cluster (N=100) and an atom, as a function of the highest order of many-body terms included. Triangles (circles) correspond to the z (x) -configuration of the monolayer cluster. Open symbols are from direct calculation of 2-body and 3-body sum. Dotted curves are drawn to guide eyes.

**FIGURE 9.** (Color online) Ratios $R_m$, $R_3$ and $R_{4h}$ for dispersion interaction between two identical cubic clusters with one side of each cluster lying perpendicular to the connecting line as shown in the inset: $R_m$ (red open square), $R_3$ (blue filled square) and $R_{4h}$ (black cross). The data of $R_m$ and $R_{4h}$ coincide within the resolution of the figure.

**FIGURE 10.** (Color online) Ratios $R_m$, $R_3$ and $R_{4h}$ for dispersion interaction between two identical linear clusters: $R_m$ (filled symbols), $R_3$ (open symbols) and $R_{4h}$ (curves) for configurations of zz (black squares and a solid curve), zx (blue circles and a dotted curve), xx (green triangles and a broken curve with single dot in between), and yx (red diamonds and a broken curve with double dots in between). Configurations are indicated in the figure on $R_{4h}$ curves.

**FIGURE 11.** (Color online) Ratios $R_m$, $R_3$ and $R_{4h}$ for dispersion interaction between two identical square prism clusters: $R_m$ (filled symbols), $R_3$ (open symbols) and $R_{4h}$ (curves) for configurations of zz (black squares and a solid curve), zx (blue circles and a dotted curve), xx (green triangles and a broken curve with single dot in between), and yx (red diamonds and a broken curve with double dots in between). Configurations are indicated in the figure on $R_{4h}$ curves.

**FIGURE 12.** (Color online) Ratios $R_m$, $R_3$ and $R_{4h}$ for dispersion interaction between two identical square monolayer clusters: $R_m$ (filled symbols), $R_3$ (open symbols) and $R_{4h}$ (curves) for configurations of xx (black triangles and a broken curve with single dot in between), xz (blue circles and a dotted curve) and zz (red squares and a solid curve). Configurations are indicated in the figure on $R_{4h}$ curves. Put labels zz and xz on the other two drawings.

**FIGURE 13.** (Color online) Fraction of total VDW interaction energy between two decamers, as a function of the highest order of many-body terms included. Symbols adopted for various configurations are squares(zz), triangles(xx), circles(xz), and diamonds(xy). Open symbols are from direct calculation of 2-body and 3-body sum. Dotted curves are drawn to guide eyes.

**FIGURE 14** (Color online) Fraction of total VDW interaction energy between two square monolayers (N=100), as a function of the highest order of many-body terms included. Symbols adopted for various orientations are triangles(xx), squares(zz) and circles(xz). Open symbols are from direct calculation of 2-body and 3-body sum. Dotted curves are drawn to guide eyes.

**FIGURE 1 [Kim et. al.]**

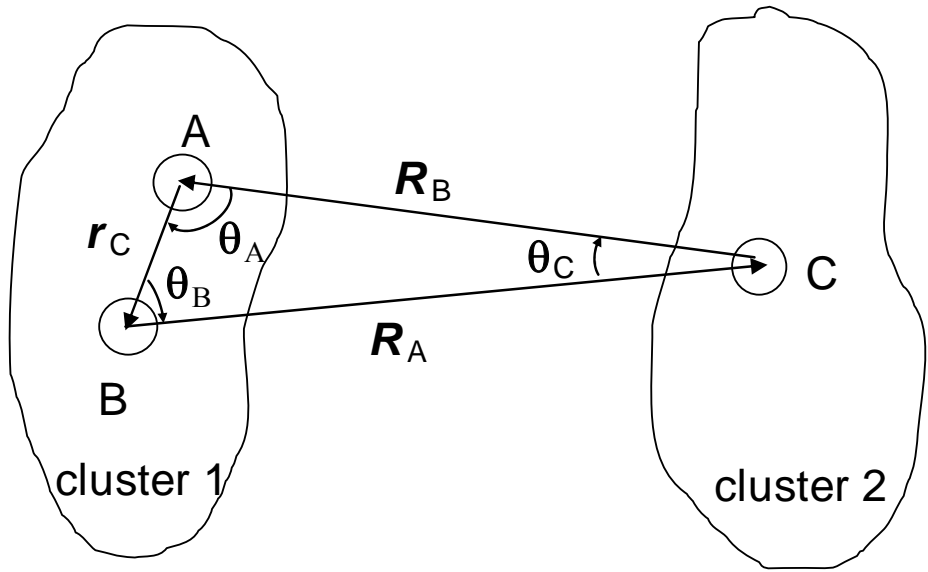

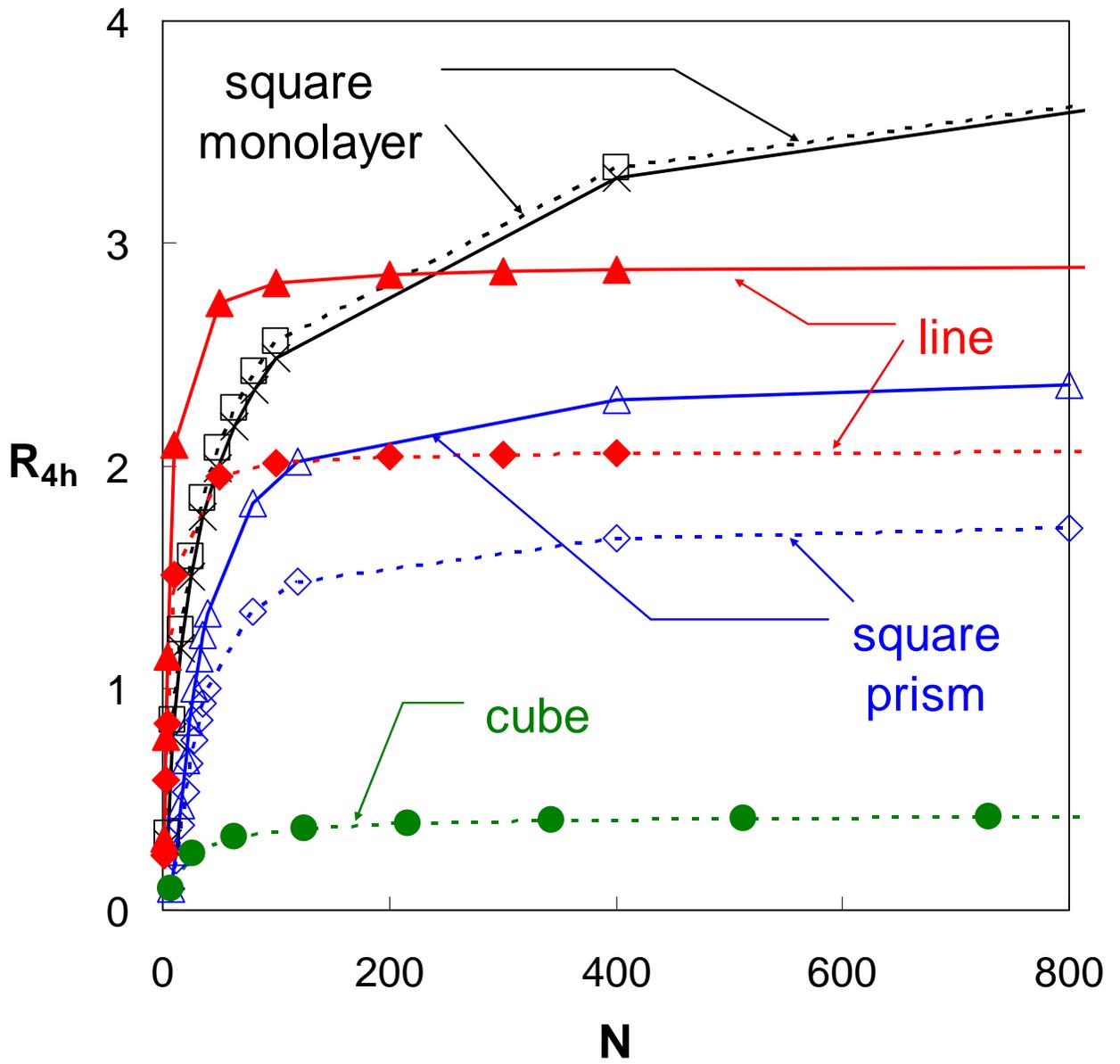

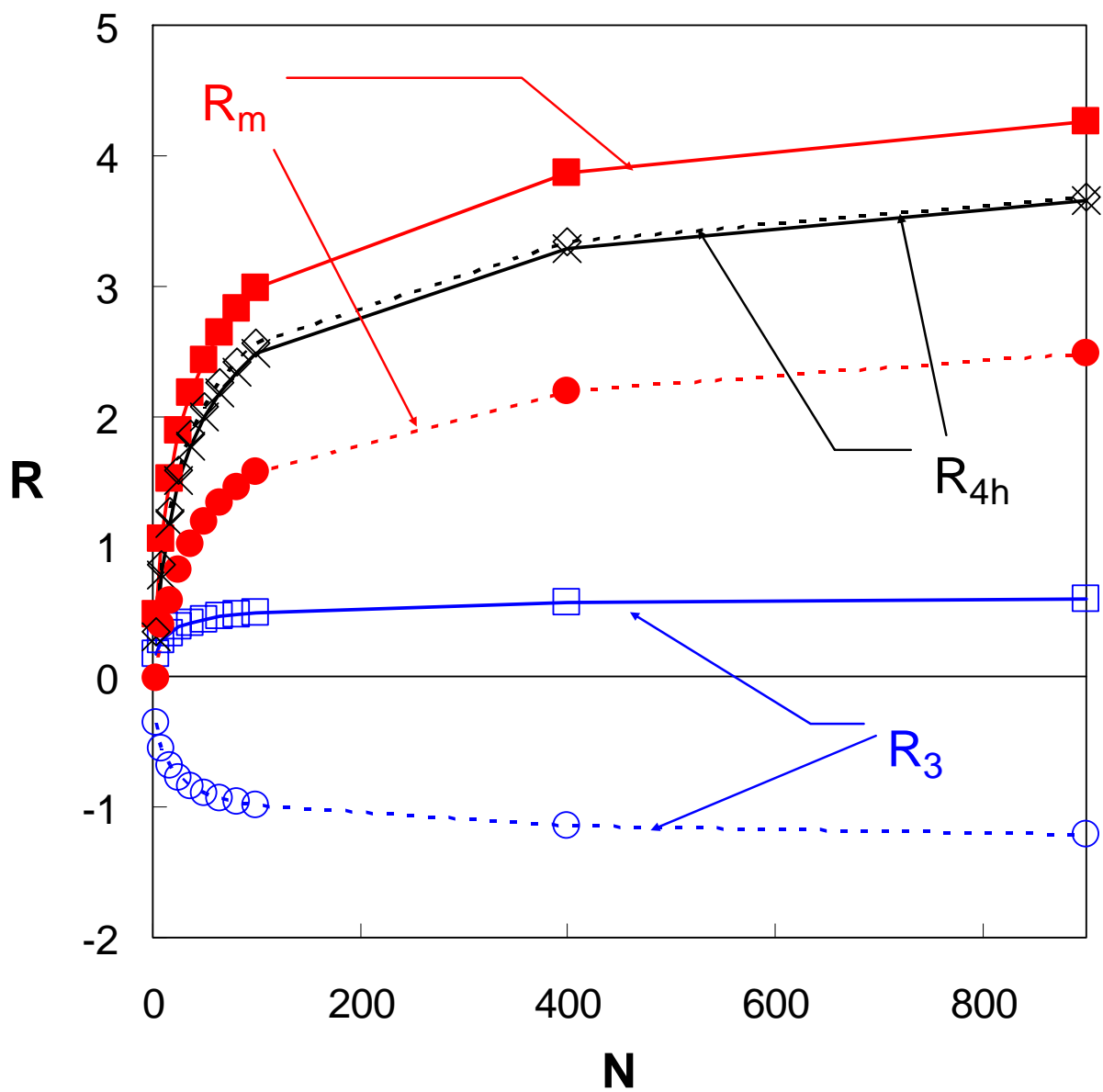

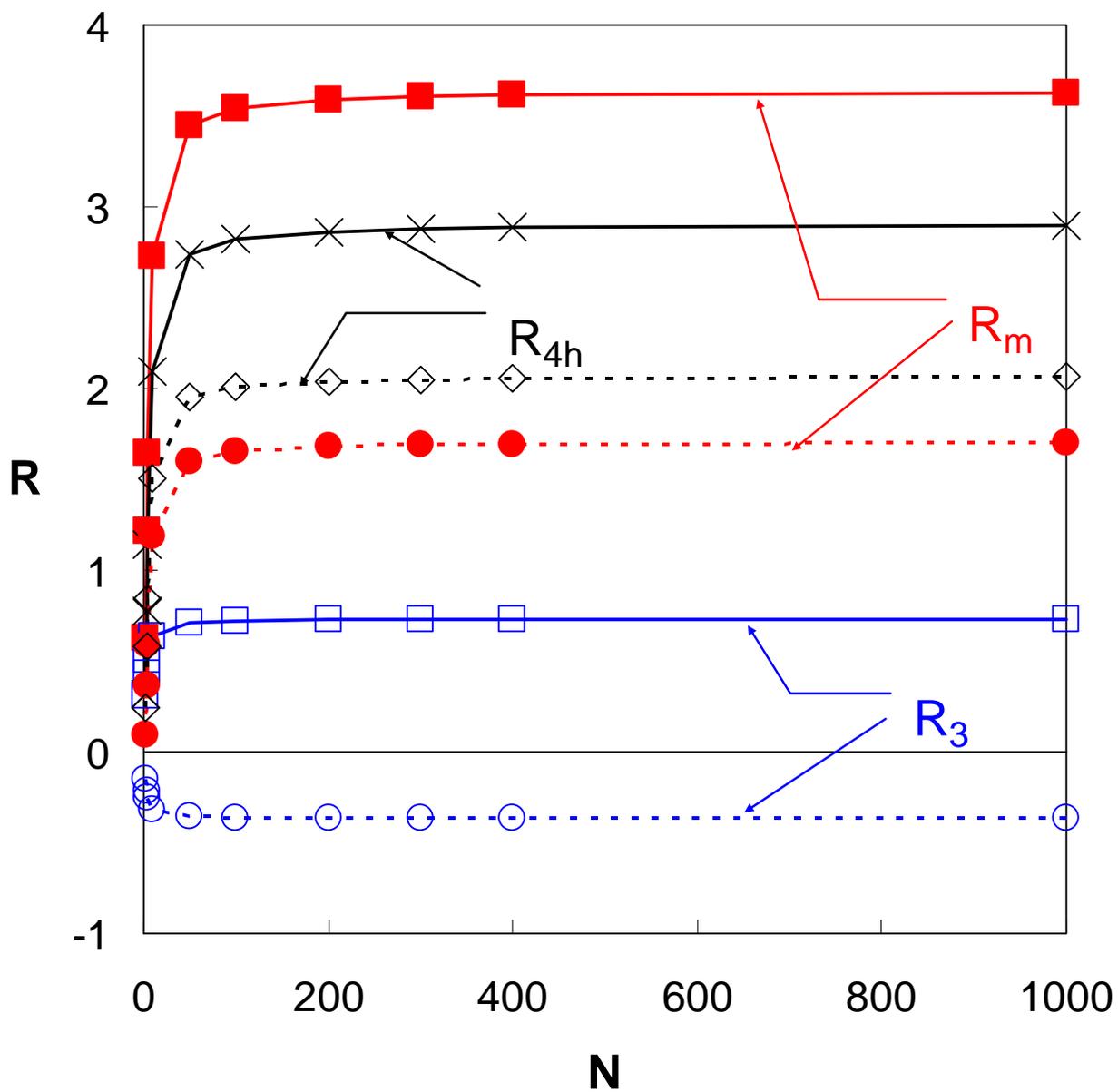

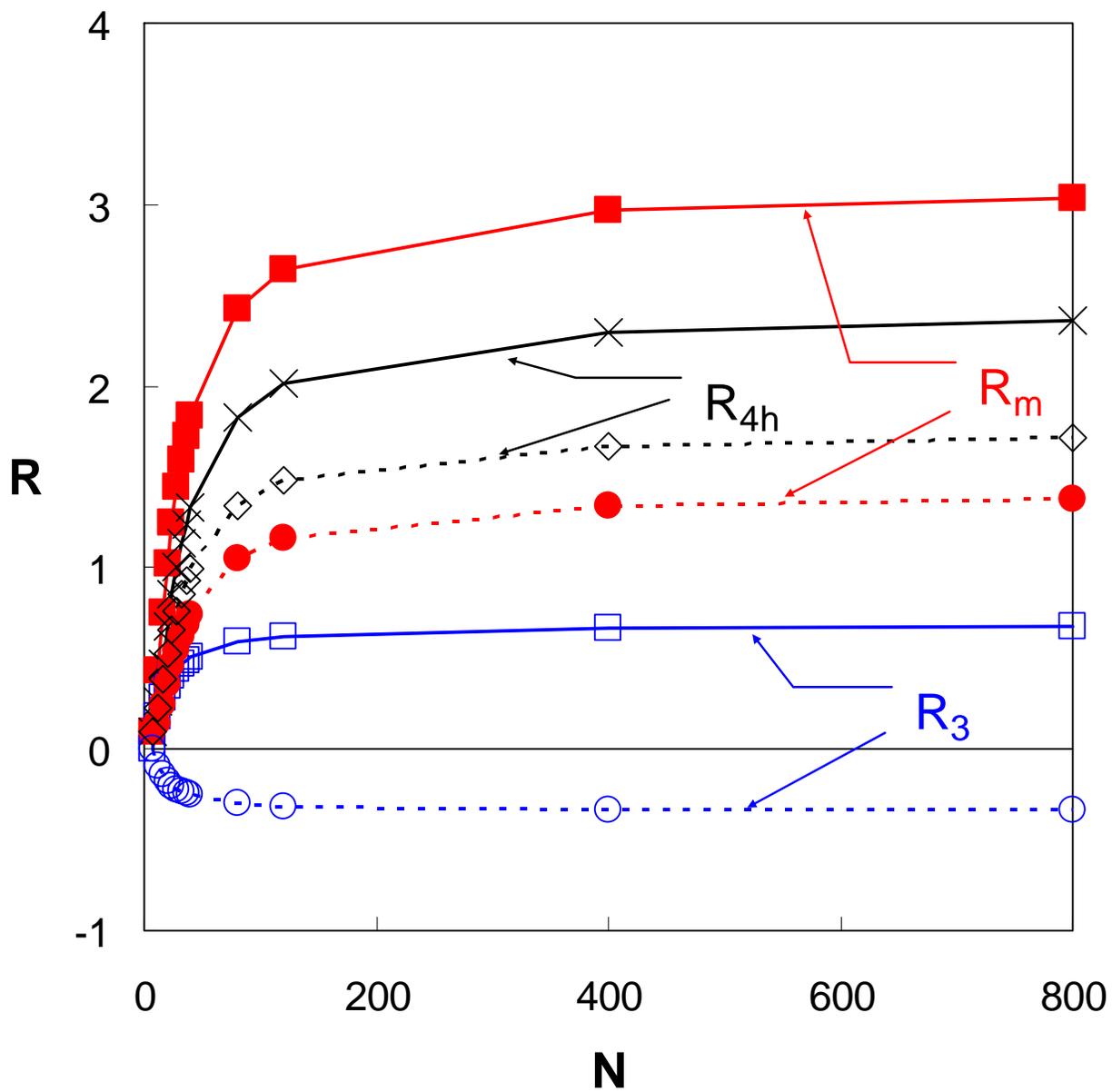

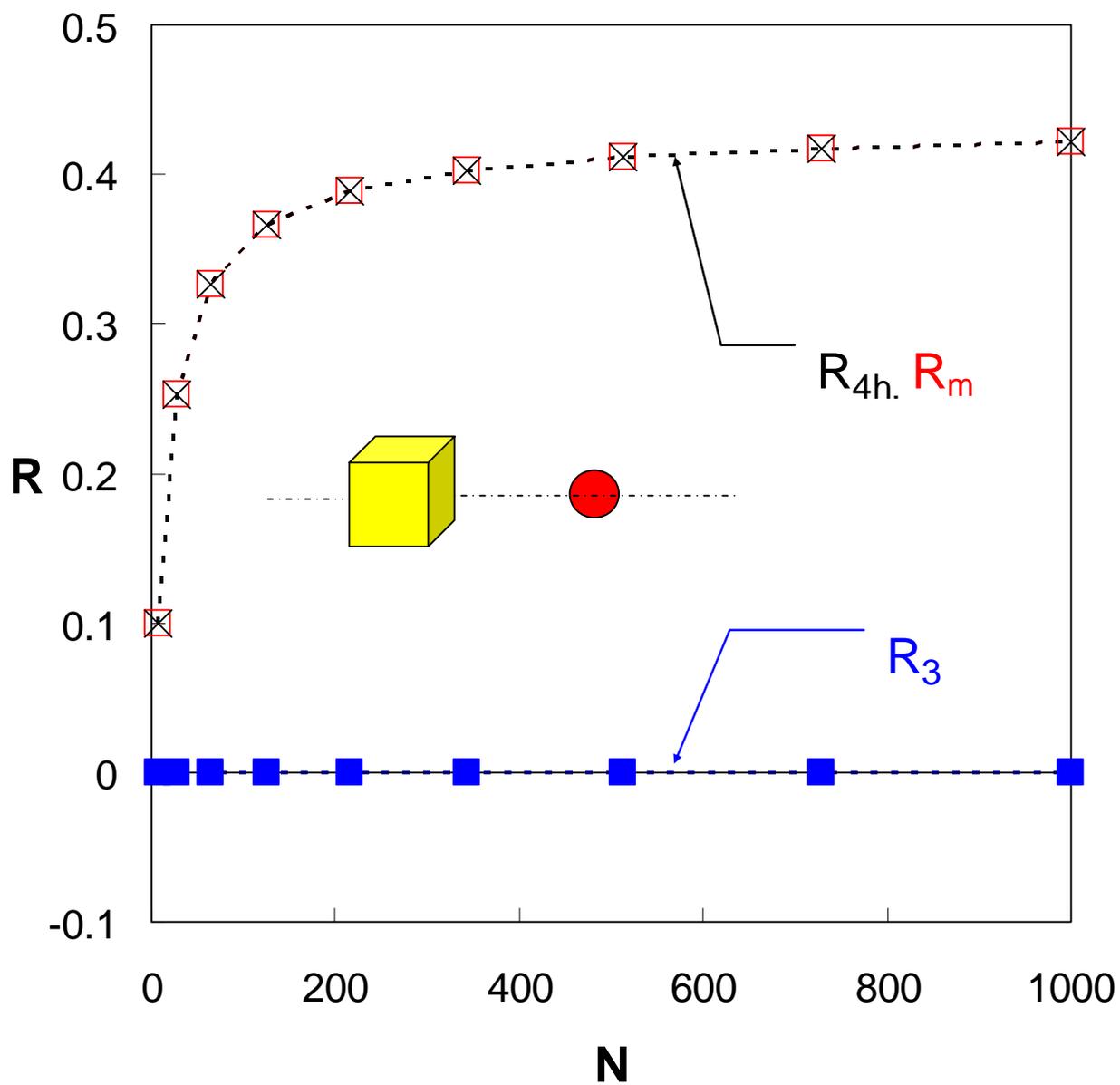

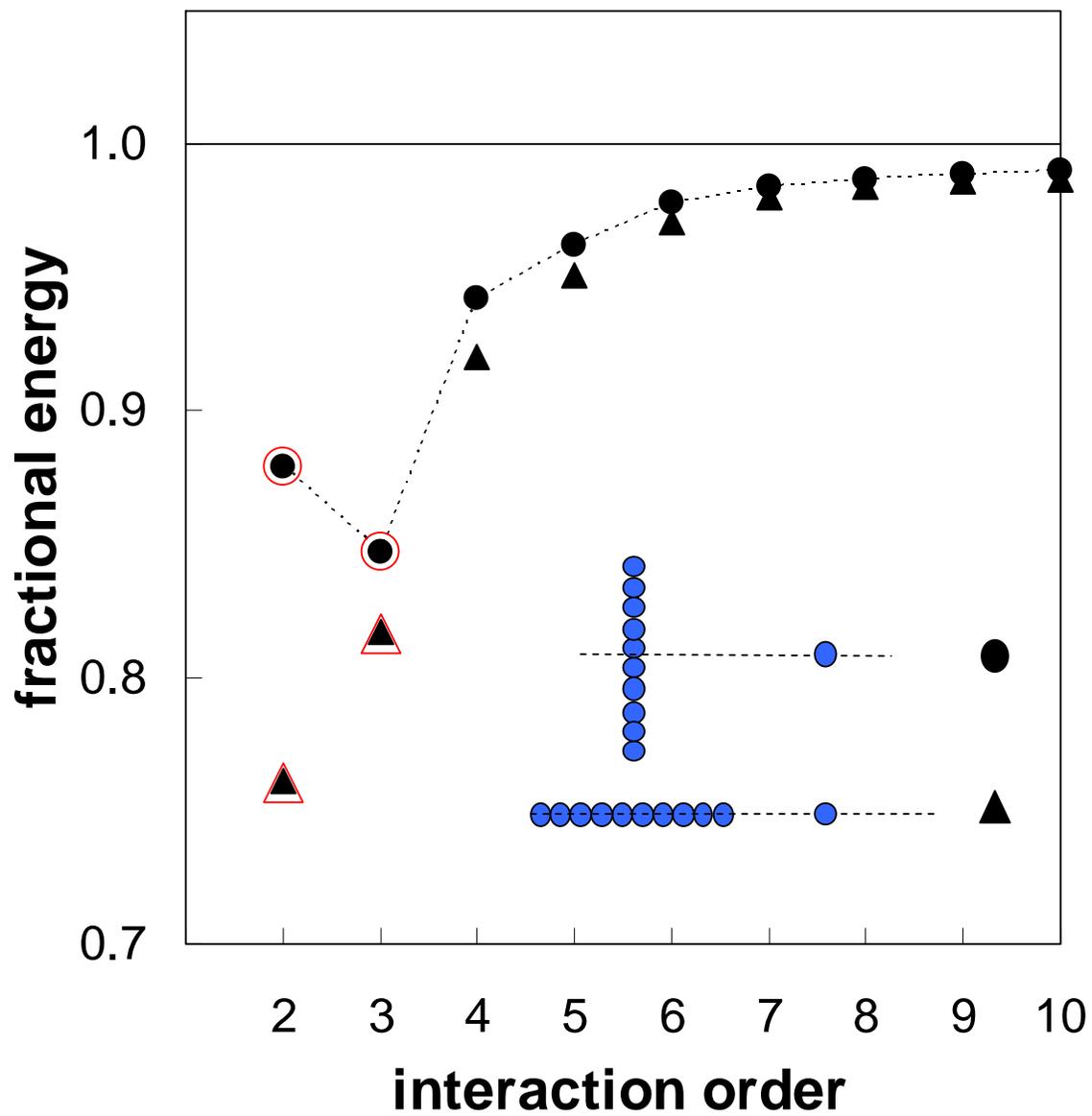

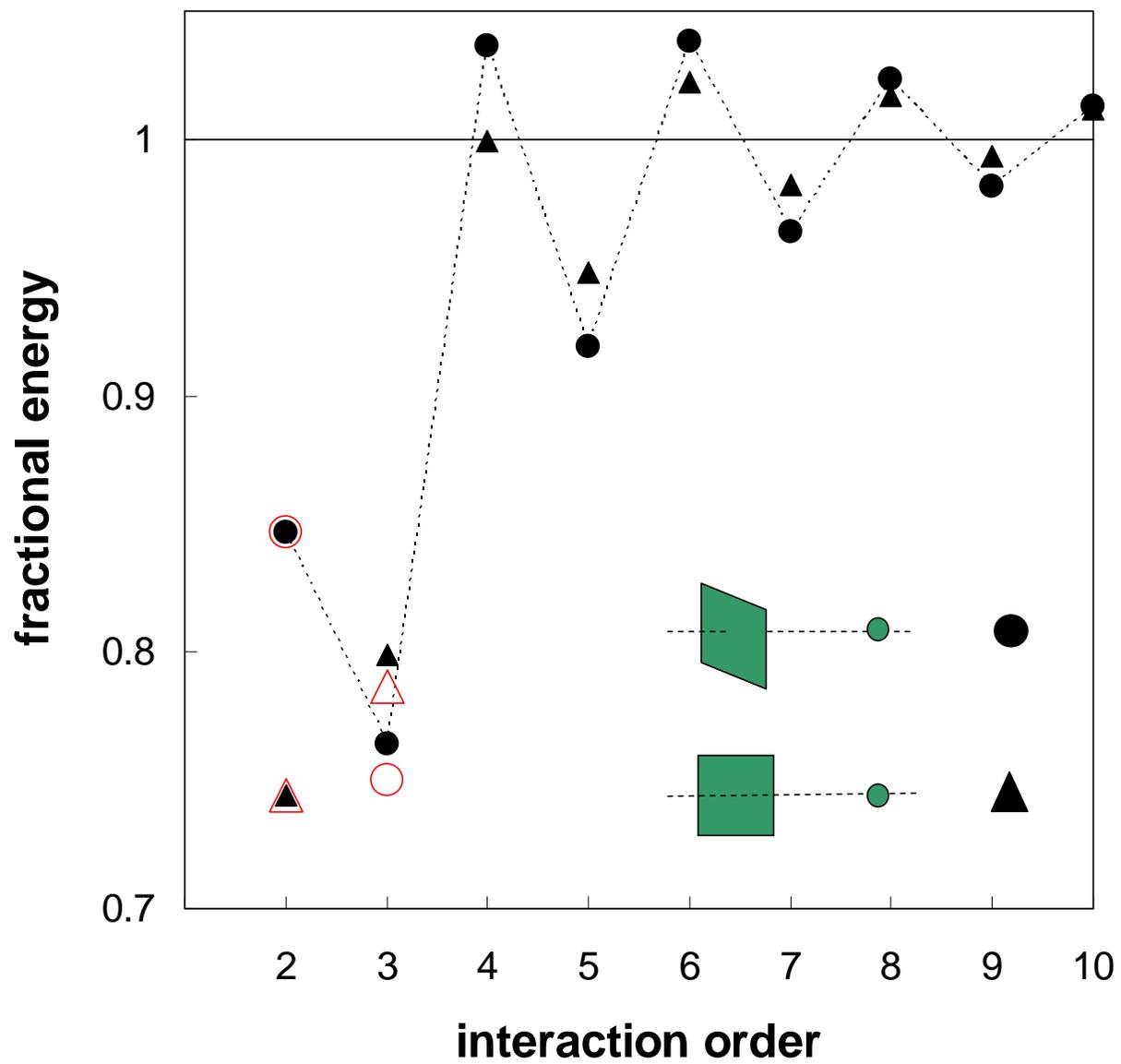

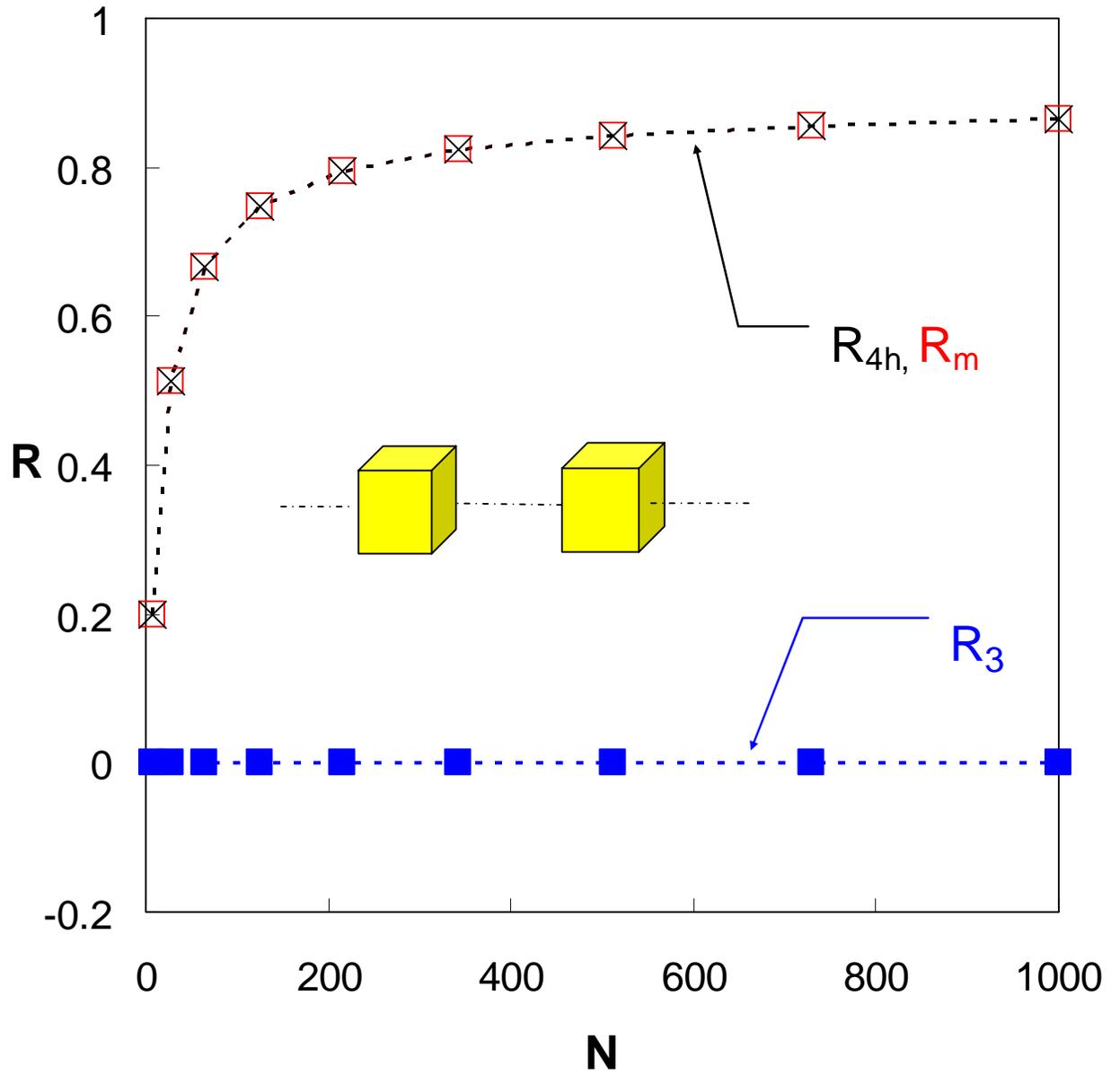

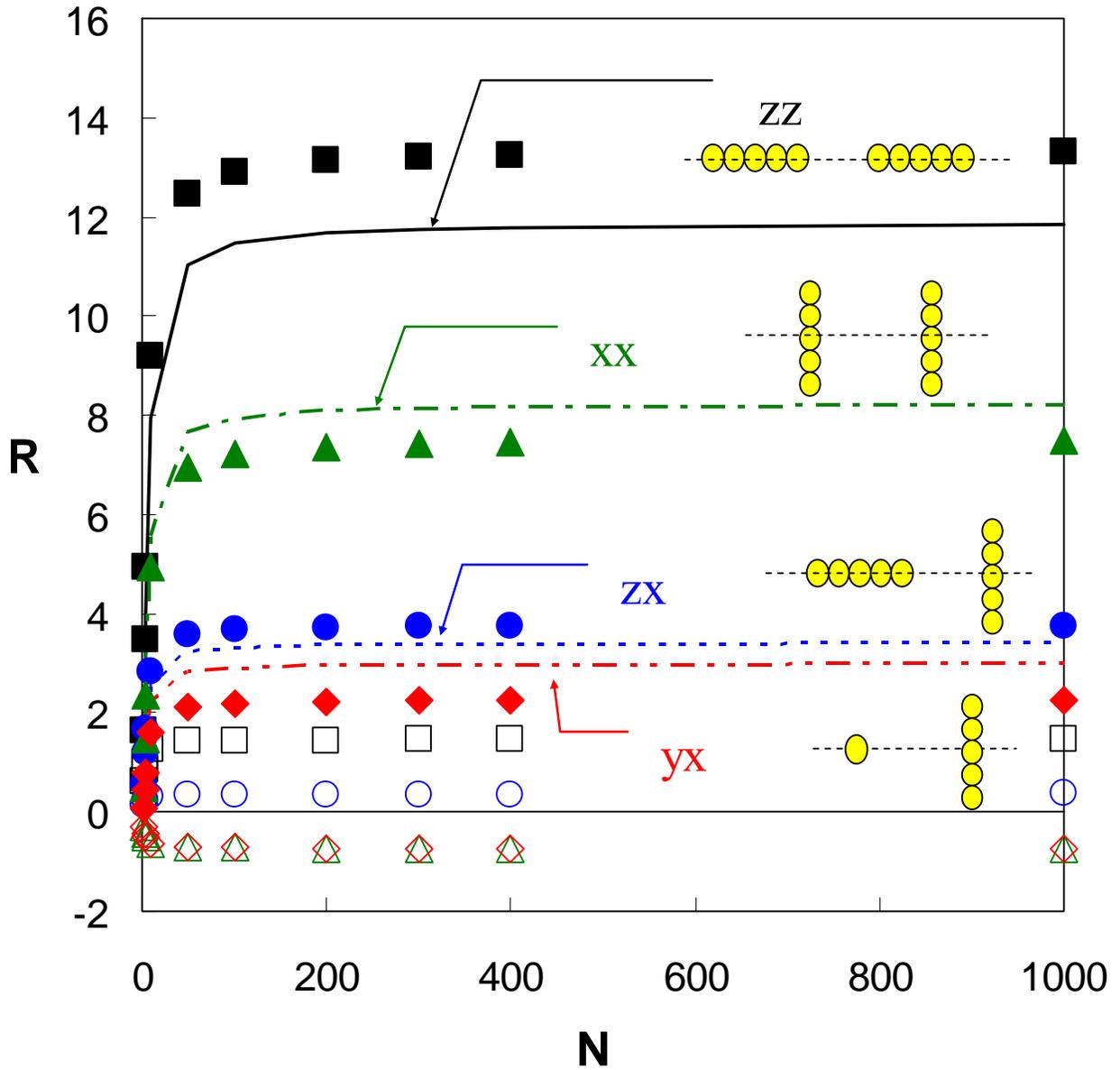

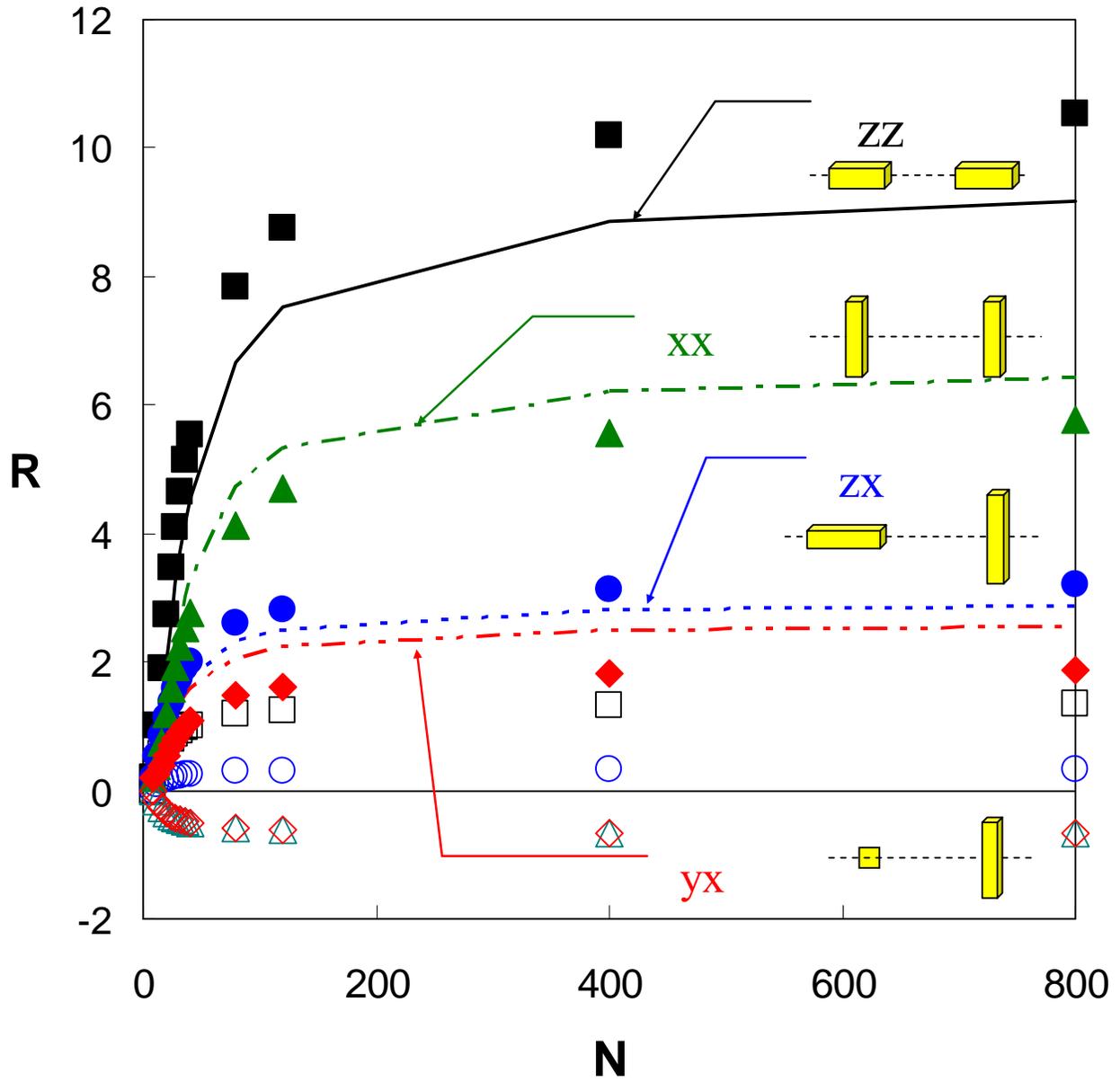

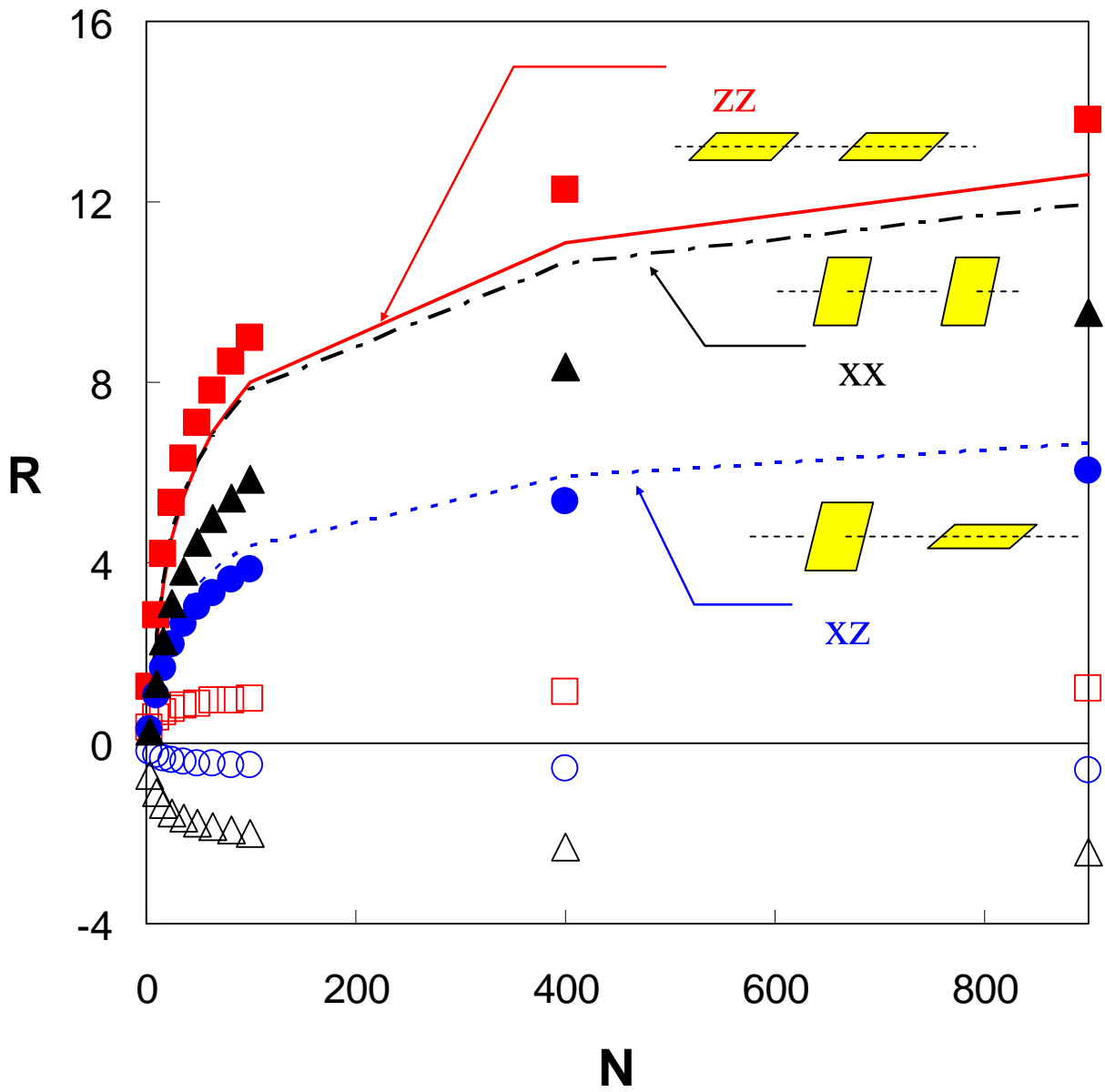

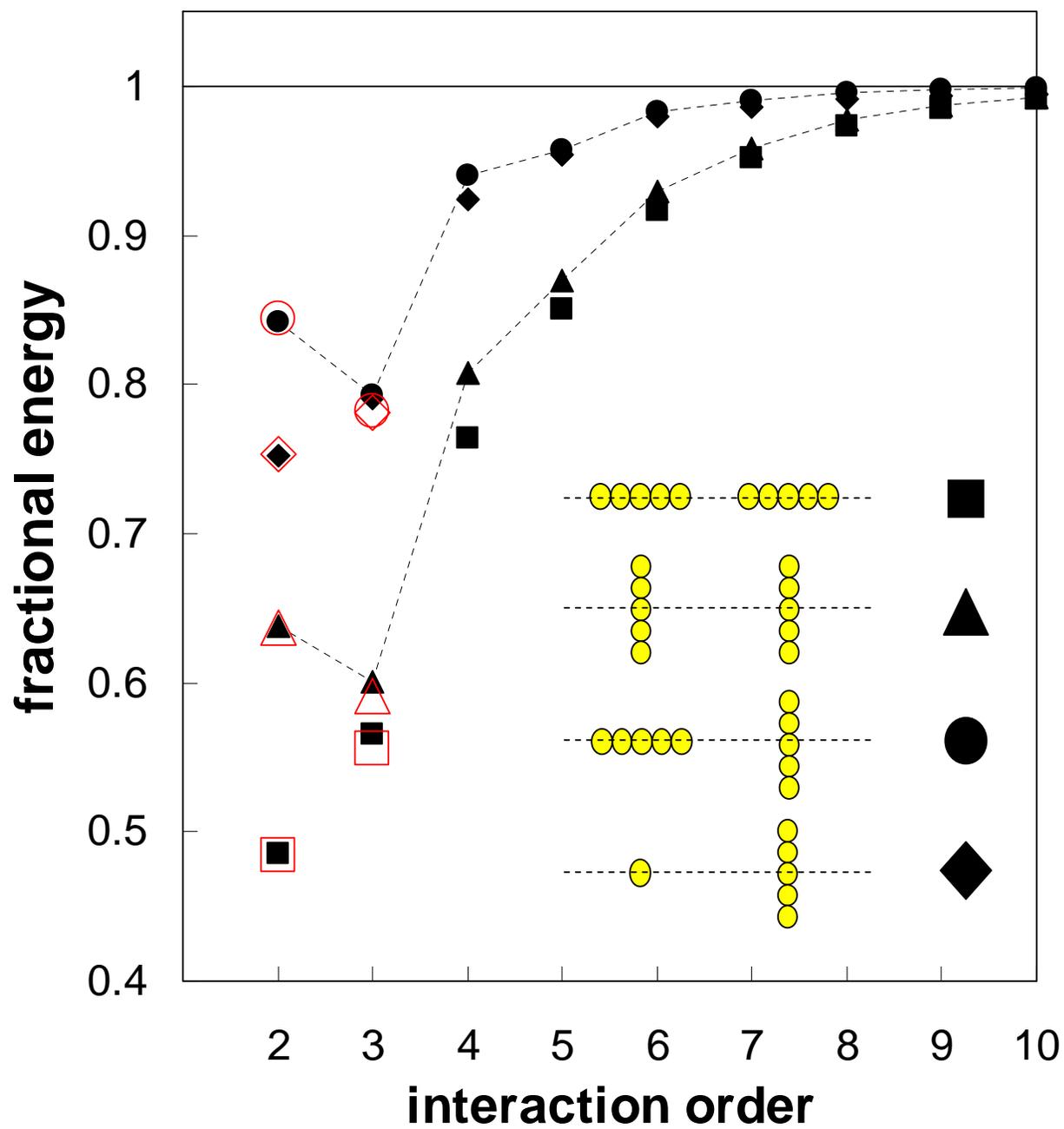

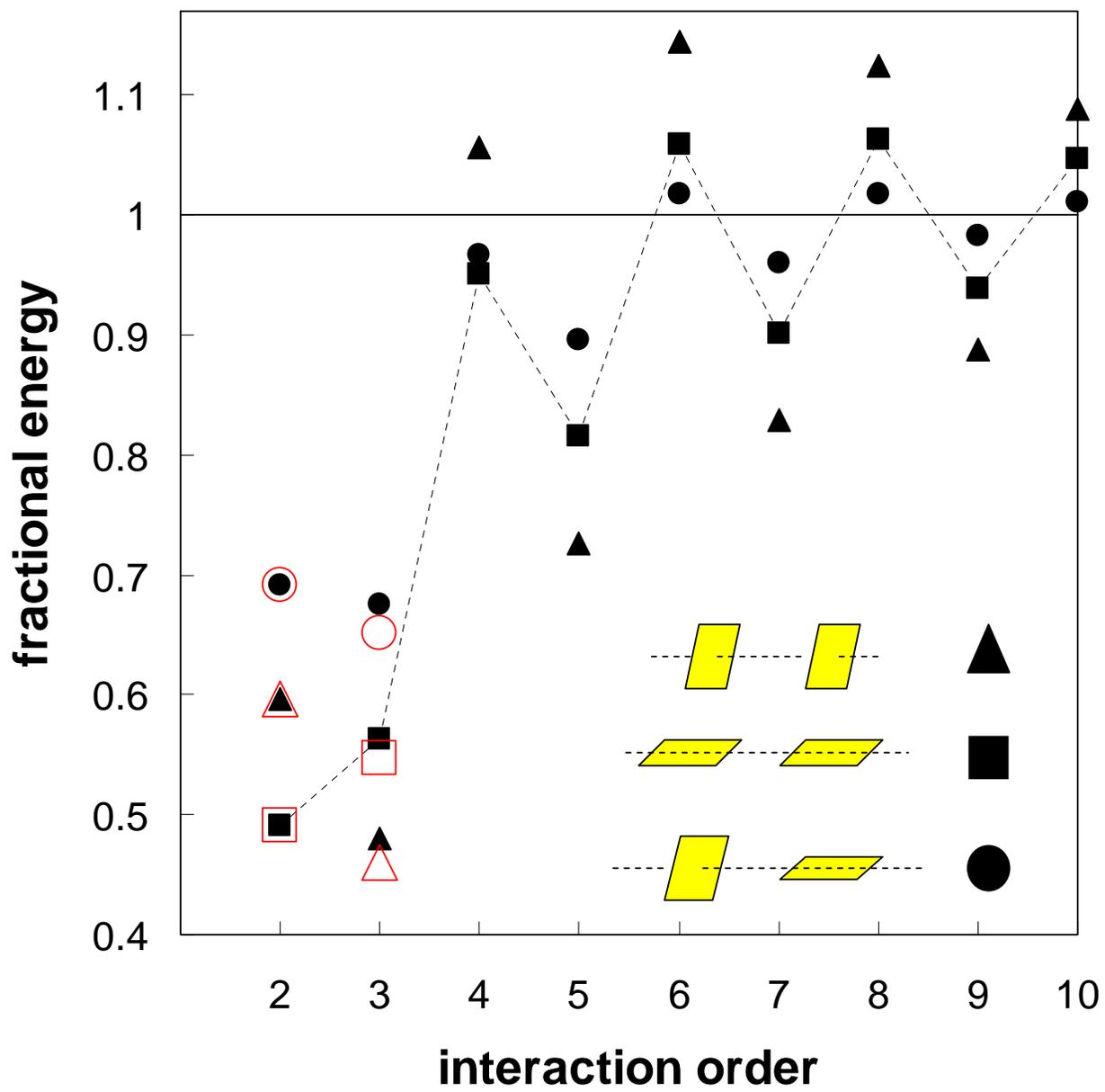